\documentclass[twocolumn,showpacs,showkeys,preprintnumbers,amsmath,amssymb,superscriptaddress,unsortedaddress]{revtex4}

\usepackage{graphicx}
\usepackage{dcolumn}
\usepackage{bm}

\begin{document}

\preprint{APS/123-QED}

\title{Structural phase transition in Tm$_{x}$Fe$_{1-x}$Se$_{0.85}$ (Tm = Mn and Cu) and its relation to superconductivity}

\author{Tzu-Wen Huang}\email{twhuang@phys.sinica.edu.tw}\affiliation{Institute of Physics, Academia Sinica, Nankang, Taipei, Taiwan}
\author{Ta-Kun Chen}\affiliation{Institute of Physics, Academia Sinica, Nankang, Taipei, Taiwan}
\author{Kuo-Wei Yeh}\affiliation{Institute of Physics, Academia Sinica, Nankang, Taipei, Taiwan}
\author{Chung-Ting Ke}\affiliation{Institute of Physics, Academia Sinica, Nankang, Taipei, Taiwan}
\author{Chi Liang Chen}\affiliation{Institute of Physics, Academia Sinica, Nankang, Taipei, Taiwan}
\author{Yi-Lin Huang}\affiliation{Institute of Physics, Academia Sinica, Nankang, Taipei, Taiwan}
\author{Fong-Chi Hsu}\affiliation{Institute of Physics, Academia Sinica, Nankang, Taipei, Taiwan}
\author{Maw-Kuen Wu}\email{mkwu@phys.sinica.edu.tw}\affiliation{Institute of Physics, Academia Sinica, Nankang, Taipei, Taiwan}
\author{Phillip M. Wu}\affiliation{Department of Physics, Duke University, Durham, NC 27708}
\author{Maxim Avdeev}\affiliation{Bragg Institute, Australian Nuclear Science and Technology Organisation, PMB 1, Menai, NSW 2234, Australia}
\author{Andrew J. Studer}\affiliation{Bragg Institute, Australian Nuclear Science and Technology Organisation, PMB 1, Menai, NSW 2234, Australia}

\date{July 19, 2009}

\begin{abstract}
In this letter, we report the results of detailed studies on Mn- and Cu-substitution to Fe-site of $\beta$-FeSe, namely Mn$_{x}$Fe$_{1-x}$Se$_{0.85}$ and Cu$_{x}$Fe$_{1-x}$Se$_{0.85}$. The results show that with only 10 at\% Cu-doping the compound becomes a Mott insulator. Detailed temperature dependent structural analyses of these Mn- and Cu-substituted compounds show that the structural transition, which is associated with the changes in the building block FeSe$_4$ tetrahedron, is essential to the occurrence of superconductivity in $\beta$-FeSe.
\end{abstract}

\maketitle

The iron-pnictide \cite{kamihara2006, kamihara2008, takahashi2008, chen2008, zhao2008, rotter2008} and $\beta$-FeSe \cite{hsu:2008oh, yeh:2008ez} superconductors have become a focus of condensed matter research in the past year. In the iron-based superconductors, there exists a structural transition at temperature (Ts) much higher than the superconducting transition point (Tc). At Ts the tetragonal lattice (P4/nmm) distorts into a lower symmetry monoclinic lattice (P112/n) (or orthorhombic with the defined a-b plane rotated about 45$^{\circ}$ with respect to the original lattice). In both the LaFeAsO (1111) and BaFe$_2$As$_2$ (122) families it was suggested that this phase transition, which is accompanied with an antiferromagnetic state developed at around the same temperature, has to be suppressed either by chemical doping or applying external pressure in order to observe superconductivity \cite{zhao2008, drew2009, cruz:2008, nomura2008, luetkens2009, kimber2009}. However, this distortion seems to be indispensable to the superconductivity in the FeSe (11) compound \cite{hsu:2008oh, yeh2008, wu2009}. Preliminary M\"{o}ssbauer measurements \cite{mcqueen2009PRB, mcqueen2009} suggested no magnetic ordering developed below Ts and beyond Tc in FeSe. Yet, the existence of short-range ordering or spin fluctuation could not be totally ruled out.

In order to further investigate the distortion issue, substitutions on Fe sites were studied earlier in 11 \cite{wu2009}, and in the 1111 \cite{matsuishi2009} and 122 families \cite{kimber2009,sefat2008,chu2009}. Among all substituent alternatives, transition metals, especially those with unpaired 3d electrons such as Mn, Co, Ni, and Cu \cite{williams2009}, would be of most interest for their comparable ionic sizes to Fe, and the potential to investigate in more detail the interplay between magnetism and superconductivity, which may also lead to better insight into the origin of superconductivity in this class of materials.

We reported earlier our preliminary results on a series of 3d-transition-metal substituted FeSe$_{1-x}$ compounds \cite{wu2009}. For all 3d-elements (from Ti to Cu) with 10 at\% substitution, we found only Mn, Co, Ni, and Cu could retain the tetragonal structure. We later decided to investigate in detail the Cu$_{x}$Fe$_{1-x}$Se$_{0.85}$ and Mn$_{x}$Fe$_{1-x}$Se$_{0.85}$ samples for comparison as we found only 3 atomic percent (at\%) Cu-doping completely suppressed superconductivity, whereas up to 30 at\% Mn-substitution only slightly decreased the superconducting transition temperature.

Cu and Mn substituted samples were prepared with method similar to that in \cite{wu2009}. TEM analysis was performed on powder samples suspended on gold grids coated with amorphous carbon in a JEOL 2100F transmission electron microscope equipped with STEM and EDX spectral analytical parts. The X-ray absorption near edge spectra were measured with calibrated standard iron foil at BL16A NSRRC with energy resolution about 0.1 eV. Cell parameters were calculated from the experiments performed in synchrotron source (BL12b2 at SPring 8 and BL13A at NSRRC) with an incident beam of wavelength 0.995 $\AA$. Neutron powder diffraction data were collected using Echidna and Wombat diffractometers\cite{liss2006,studer2006} at the OPAL reactor, Australia. The samples were loaded in 6 mm cylindrical vanadium cans and data were collected in the temperature range 3-300 K using wavelength of 1.885 $\AA$. The resistance measurements were carried out using the standard 4-probe method with silver paste for contacts.

\begin{figure}
\includegraphics[width=0.4\textwidth]{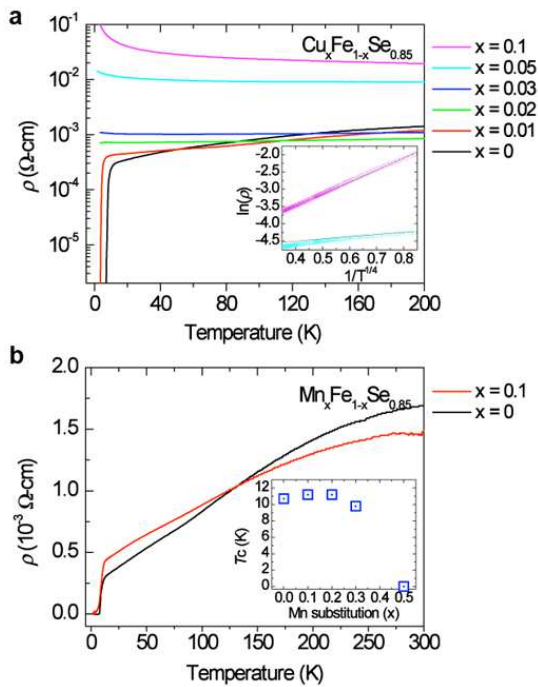}
\caption{\label{fig1} (a) Temperature-dependent electrical resistivity ($\rho$) at zero magnetic field for bulk Cu$_{x}$Fe$_{1-x}$Se$_{0.85}$ (x=0, 0.01, 0.02, 0.03, 0.05 and 0.1) samples. Inset shows the low temperature resistivity plotted against 1/T$^{1/4}$ (b) Temperature dependence of electrical resistivity for bulk Mn$_{x}$Fe$_{1-x}$Se$_{0.85}$ (x=0 and 0.1) samples. Inset shows Tc as a function of Mn substitution.}
\end{figure}

\begin{figure}
\includegraphics[width=0.4\textwidth]{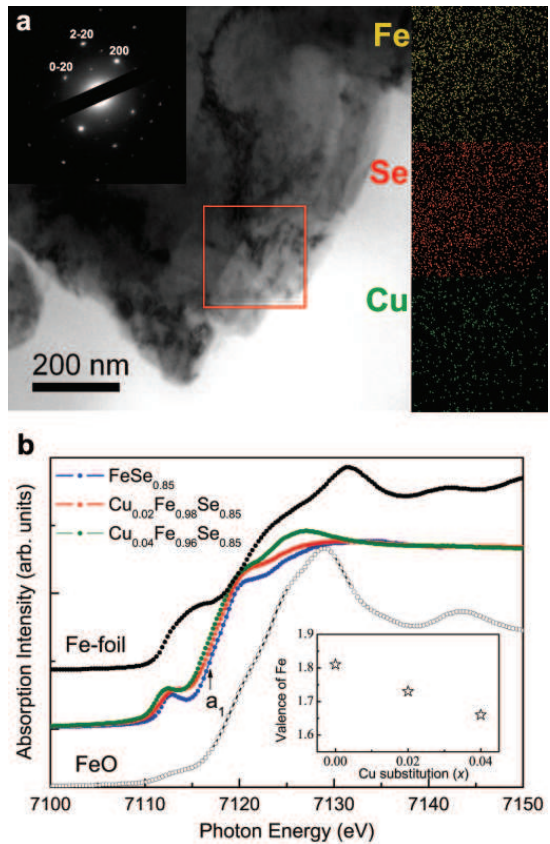}
\caption{\label{fig2} (a) The bright field TEM image and the corresponding selected-area electron diffraction of Cu$_{0.04}$Fe$_{0.96}$Se$_{0.85}$ powder sample, aligned along the (001) direction revealing the 4-fold symmetry of the tetragonal structure. At the right side we present the STEM/EDX elemental mappings of the area of interest marked by a red square using a 2-$\AA$ electron-probe, which demonstrate the random distribution of copper in the sample. (b) The x-ray absorption near-edge structure (XANES) in Fe K-edge for Cu$_{x}$Fe$_{1-x}$Se$_{0.85}$ (x=0-0.04), Fe foil (upper), and FeO (bottom). The inset plots the valence states of Fe, which are calculated from the first derivative of the XANES spectra.}
\end{figure}

\begin{figure*}
\includegraphics[width=0.8\textwidth]{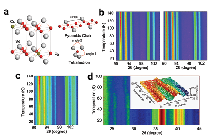}
\caption{\label{fig3} (a) Crystal structure of Cu$_{x}$Fe$_{1-x}$Se$_{0.85}$, sketched schematically with Fe in red, Cu in dark yellow and Se in grey colour. The pyramids chain and tetrahedron with respect to iron are shown to the right. (b) Temperature dependence of NPD for Cu$_{0.01}$Fe$_{0.9}$Se$_{0.85}$ (left) and Cu$_{0.1}$Fe$_{0.9}$Se$_{0.85}$ (right) bulk samples. (c) NPD of Mn$_{0.1}$Fe$_{0.9}$Se$_{0.85}$ bulk sample. Similar structural change is observed as evidenced by the peak splitting in (220), (221) and (114) reflections at $\sim$85 K. (d) NPD for Mn$_{0.1}$Fe$_{0.9}$Se$_{0.85}$ at low q range from 100 to 10 K.}
\end{figure*}


Figure \ref{fig1} shows the temperature dependence of electrical resistivity of Cu$_{x}$Fe$_{1-x}$Se$_{0.85}$ and Mn$_{x}$Fe$_{1-x}$Se$_{0.85}$ compounds with various x values. Superconducting transition in Cu$_x$Fe$_{1-x}$Se$_{0.85}$ (Fig. \ref{fig1}a) was observed only in samples with x $\leq$ 0.02. For x $\geq$ 0.03, the compound gradually becomes semiconductor-like \cite{williams2009}. Detailed analysis of the temperature dependence of resistivity shows that for 10 at\% Cu-doping sample the resistivity, as shown in the inset of Fig. 1a, fits well with the 3D Mott variable range hopping transport. In contrast, Mn$_{x}$Fe$_{1-x}$Se$_{0.85}$ (Fig. \ref{fig1}b) remains metallic and superconducting for x as high as 0.3 with only very little variation in Tc, as shown in the inset of Fig. \ref{fig1}b.

It was surprising that only 3 at\% Cu-doping makes the sample become an insulator. Figure \ref{fig2}a shows a TEM image of a Cu$_{0.04}$Fe$_{0.96}$Se$_{0.85}$ powder sample aligned with the c-axis, in a way that the Fe-Fe plane or the Se-Se plane is parallel to plane of this page. The selected-area electron diffraction shows that, the reflections at the (hkl), h+k=2n, h=k=odd positions are quite strong, while they are expected to be very weak in FeSe. This strongly suggests the successful substitution of Cu into Fe site. The STEM/EDX elemental mappings (right panel of Fig. \ref{fig2}a) of the area of interest marked by a red square demonstrate no particular feature of copper in the sample suggesting homogeneous dispersion of Cu over the whole sample. The 2-$\AA$ scanning electron-probe, which is smaller than the Fe-Fe or Se-Se distance, is expected to be able to resolve any clustering or non-uniformity in the samples. The Fe and Se concentrations are as well found uniformly distributed in the sample.

The XAS Fe K-edge spectra are shown in Fig. \ref{fig2}b, which are normalized at the photon energy $\sim$100 eV from the absorption edge at $E_0$=7112 eV (pure Fe). The feature marked as a1 is mainly due to the transition from Fe 1s to the 4sp state as in the FeSe$_x$ series \cite{chen2009unpub}. A comparison of the spectra of the standard (Fe and FeO) and the Cu$_{x}$Fe$_{1-x}$Se$_{0.85}$ samples with x=0-0.04 reveals the energy shifting at the a1 regions around 7116.8 eV with increasing x value. The results indicate that the variation of Fe valence, which is shown in the inset of Fig. \ref{fig2}b, decreases from +1.81 at x=0 to +1.66 at x=0.04. The linear shifting of absorption edge gives additional support to the random distribution of Cu over Fe, since inhomogeneity may give rise to deviations in the absorption spectra. In the Fe 4p states \cite{kisiel1999}, a smooth feature from 8 to 15 eV above the edge also showed a tendency towards larger areas, suggesting a systematic increase of unoccupied states above the Fermi level as more Cu is substituted. These results may provide a rational explanation to the observed resistivity increases, and eventual insulating behavior, in higher Cu-doping samples.

In Fig. \ref{fig3} we show the schematic crystal structure of $\beta$-FeSe and the temperature-dependent neutron powder diffraction patterns (NPD) of Cu- and Mn-substituted samples. Figure \ref{fig3}b is the temperature dependence of neutron scattering for Cu$_{0.01}$Fe$_{0.9}$Se$_{0.85}$ (left) and Cu$_{0.1}$Fe$_{0.9}$Se$_{0.85}$ (right) bulk samples. Peak splitting was observed in (220), (221) and (114) reflections at $\sim$60 K in the Cu$_{0.01}$Fe$_{0.9}$Se$_{0.85}$ sample. However, no splitting could be identified for any peak in Cu$_{0.1}$Fe$_{0.9}$Se$_{0.85}$ sample from 140 to 10 K, indicating the absence of any structural distortion in the 10 at\% Cu-doping samples. On the other hand, in the NPD of Mn$_{0.1}$Fe$_{0.9}$Se$_{0.85}$ bulk sample, Fig. \ref{fig3}c, peak splitting is observed in (220), (221) and (114) reflections at $\sim$85 K indicating the onset of structural phase transition. This phase transition could be described by a structural distortion from tetragonal lattice (P4/nmm) to monoclinic (P112/n), which is much the same as observed in the FeSe \cite{hsu:2008oh,mok2009} and FeSe$_{0.5}$Te$_{0.5}$ at temperatures below $\sim$100 K \cite{yeh:2008ez}. Moreover, if viewing along the (110) direction, the lattice that distorts from tetragonal to monoclinic does not destroy the magnetic symmetry, allowing superconductivity to occur\cite{yeh:2008ez}. In the neutron data for Mn$_{0.1}$Fe$_{0.9}$Se$_{0.85}$ from 100 to 10 K, as shown in Fig. \ref{fig3}d, we observed several Bragg peaks at low q range suggesting incommensurate magnetic ordering at q=1.12, 1.33, 1.43, 1.61 and 1.92 $\AA^{-1}$ (where $q=4\pi sin(\theta)/\lambda$). The inset shows the intensity of magnetic reflections in log scale at 25.82$^{\circ}$ and 28.72$^{\circ}$ (q=1.43 $\AA^{-1}$ and 1.61 $\AA^{-1}$) and indicates the magnetic transition at 75 K in Mn$_{0.1}$Fe$_{0.9}$Se$_{0.85}$. The details of these observed features are currently under intensive investigation.

\begin{figure}
\includegraphics[width=0.5\textwidth]{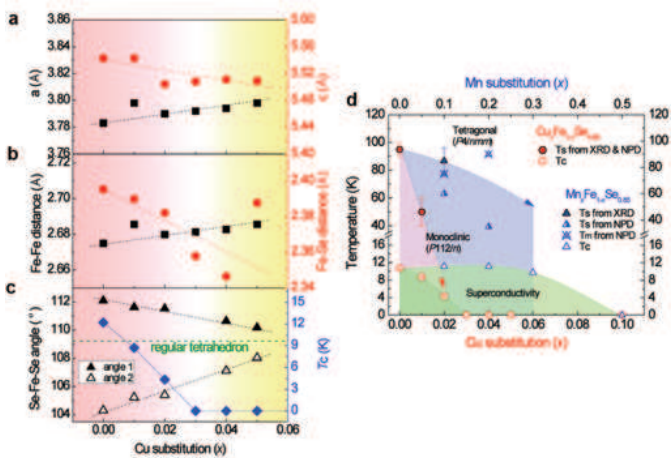}
\caption{\label{fig4} (a)-(c) Lattice constants, Fe-Se and Fe-Fe bond lengths, Se-Fe-Se bond angles and Tc of Cu$_{x}$Fe$_{1-x}$Se$_{0.85}$ as a function of Cu substitution, x. (d) The structural phase diagram of Cu$_{x}$Fe$_{1-x}$Se$_{0.85}$ and Mn$_{x}$Fe$_{1-x}$Se$_{0.85}$ determined from neutron and synchrotron X-ray powder diffraction data, and resistivity data. The solid triangles and circles indicate the onset of tetragonal to monoclinic structural distortion, Ts, and the hollow triangles and circles designate the onset of superconductivity, Tc.}
\end{figure}

Detailed Rietveld refinements of the diffraction data gives insight to the Cu substitution effect on the crystal structure of Cu$_{x}$Fe$_{1-x}$Se$_{0.85}$ (x=0-0.05). The lattice constants a and c were found slightly modified by Cu substitution, as shown in Fig. \ref{fig4}a. If seen with the tetrahedron shown in Fig. \ref{fig3}a, we found that this modification causes shrinkage in the Fe-Se bond length and slight expansion in Fe-Fe bond length (Fig. \ref{fig4}b), which is accompanied with changes in Se-Fe-Se bond angles (Fig. \ref{fig4}c). The effects combined leads toward a regular tetrahedron ($\gamma$=109.28$^{\circ}$), i.e., compression of the tetrahedron. This hardened bond strength could inhibit the structural transition at low temperature. Thus, the low temperature structural transition (Ts) was drastically suppressed and eventually disappeared when the concentration of Cu substitution exceeded 3 at\%.

Our experimental observations can be summarized in the structural phase diagram as shown in Fig. \ref{fig4}d for Cu$_{x}$Fe$_{1-x}$Se$_{0.85}$ and Mn$_{x}$Fe$_{1-x}$Se$_{0.85}$. The substitution by Cu or Mn on Fe site clearly drives down the structural transition temperature Ts, and it also reveals the correlation between Ts and Tc. As Ts deceases with increasing Cu or Mn substitution, the superconducting state is gradually suppressed. It indicates that the driving force to the formation of low temperature phase, the monoclinic P112/n structure, could be the key for the formation of superconductivity in this type of superconductors.

It is worth noting that the low-temperature structural distortion is completed by elongation along the (110) direction of tetragonal cell, which is shown by an arrow in Fig. \ref{fig3}a, revealing an one-dimension like pyramid chain through Se sites. It is natural to consider that Fermi surface nesting along the (110) direction could be mediated with this anisotropic chain. In this regard, the Fermi surface nesting along with phonon softening at proper temperatures could be an important driving force for the structural distortion. Further measurements on single crystals should be conducted before making any definite conclusion.

In summary we report the strong suppression of superconductivity by Cu substitution in the FeSe system. In comparison with Mn substitution, we found that the inhibited tetragonal to monoclinic structural phase transition should be responsible for the suppression of superconducting transition. Samples with Cu substitution over 3 at\% show no structural distortion and no superconductivity down to 2 K. Detailed structural analyses suggest that for the FeSe system the modification of FeSe$_4$ tetrahedron could be essential to the structural phase transition and thus to the origin of superconductivity.
\begin{acknowledgments}
The authors acknowledge Dr. H.C. Su and Dr. C.H. Lee of the National Tsing-Hua University for their support on the neutron experiments. This work was partially supported by the National Science Council of Taiwan. We also acknowledge the National Synchrotron Research Radiation Centre in Taiwan and the US AFOSR/AOARD for their financial support.
\end{acknowledgments}


\bibliography{cufeseprl1}

\end{document}